\documentclass[12pt,a4paper]{article}
\usepackage{amssymb,amsmath,stmaryrd,amsthm}
\usepackage[mathscr]{eucal}
\usepackage{pstricks,pst-node,pst-text,pst-3d}

\usepackage{times}

\newtheorem{lemma}{Lemma}[section]
\newtheorem{corollary}{Corollary}[section]

\newtheorem{theorem}{Theorem}[section]
\newtheorem{example}{Example}[section]

\newtheorem{remark}{Remark}[section]

\newcommand{\D}{\displaystyle}

\def\C{{\mathbb C}}
\def\F{{\mathbb F}}
\def\H{{\mathbb H}}

\def\R{{\mathbb R}}

\def\S{{\mathbb S}}

\def\Pf{{\it Proof.$\;$}}

\def\qed{\hfill $\blacksquare$}

\def\tr{\mbox{\rm tr}}

\def\({\langle}
\def\){\rangle}
\def\mb{\boldsymbol}
\def\im{{\rm i}}

\def\cA{{\mathcal A}}

\def\cC{{\mathcal C}}

\def\cS{{\mathcal S}}

\def\1{\mb1}

\def\v0{{\bf 0}}

\def\ov{\overline}

\begin{document}

\title{Game Theoretic Interaction and Decision:\\ A Quantum Analysis}

\author{Ulrich FAIGLE\thanks{Mathematisches Institut, Universit\"at zu K\"oln,
        Weyertal 80, 50931 K\"oln, Germany. Email:
        \texttt{faigle@zpr.uni-koeln.de}}         \and
        Michel GRABISCH\thanks{Corresponding author. Paris School of Economics,
          University of Paris I,
         106-112, Bd. de l'H\^opital, 75013 Paris, France. Tel. (33)
         144-07-82-85, Fax (33)-144-07-83-01. Email:
         \texttt{michel.grabisch@univ-paris1.fr}}
}

\date{\today}

\maketitle

\begin{abstract}\noindent
An interaction system has a finite set of agents that interact pairwise, depending on the current state of the system. Symmetric decomposition of the matrix of inter\-action coefficients yields the representation of states by self-adjoint matrices and hence a spectral representation. As a result, cooperation systems, decision systems and quantum systems all become visible as manifestations of special interaction systems. The treatment of the theory is purely mathematical and does not require any special knowledge of physics. It is shown how standard notions in cooperative game theory arise naturally in this context. In particular, Fourier transformation of cooperative games becomes meaningful. Moreover, quantum games fall into this framework. Finally, a theory of Markov evolution of interaction states is presented that generalizes classical homogeneous Markov chains to the present context.
\end{abstract}



\section{Introduction}\label{sec:Introduction}
In an \emph{interaction system}, economic (or general physical) agents inter\-act pairwise, but do not necessarily cooperate towards a common goal. Yet, this model arises as a natural generalization of the model of cooperative TU-games, for which already Owen~\cite{owe72} introduced the co-value as an assessment of the pairwise interaction of two cooperating players\footnote{see also Grabisch and Roubens~\cite{grro99} for a general approach}. In fact, it turns out that the context of interaction systems provides an appropriate framework for the analysis of cooperative games (\emph{cf.} Faigle and Grabisch~\cite{FaigleGrabisch16}). It is therefore of interest to investigate interaction systems in their own right.

A second motivation comes from strong arguments (put forth most prominently by Penrose~\cite{penros94}) that the human mind with its decisions and actions constitutes a physical quantum system and should therefore be modelled according to the laws of quantum mechanics. This idea has furthered a seemingly new and rapidly growing branch of game theory where in so-called \emph{quantum games} the players are assumed to execute actions on quantum systems according to the pertinent laws of physics\footnote{see, \emph{e.g.}, the surveys Grabbe~\cite{grabbe05} and Huo {\it et al.}~\cite{guo-et-al08}}. The actions of the quantum players typically transform quantum bit vectors according to the same mechanisms attributed to so-called \emph{quantum computers}. Many classical games of economic or behavioral interest have been studied in this setting. The Prisoners' Dilemma, for example, has been found to offer a Pareto optimal Nash equilibrium in the quantum model where no such equilibrium exists in the classical variant (Eisert {\it et al.}~\cite{eisert-et-al99}).

The quantum games discussed in the literature are generally
non-cooperative. Although cooperative game theory has been applied to quantum
theory\footnote{{\it e.g.}, Vourdas~\cite{vou2016}}, a co\-operative quantum
game theory has not really developped\footnote{Iqbal and
  Toor~\cite{iqbal-toor02} discuss a 3 player situation}. In general, it is felt
that quantum game theory has a highly physical side to it\footnote{see,
  \emph{e.g.}, Levy~\cite{levy17} } and that there is a gap between physics and rational game theoretic behavior\footnote{\emph{cf.} Wolpert~\cite{wolpert06}}.

\medskip
Therefore, it may be surprising that the mathematical analysis of interaction systems exhibits \emph{no} conceptual gap to exist between the ''classical'' and the ''quantum'' world in game theory. In particular, no special knowledge of quantum mechanics (or physics in general) is needed to understand game theoretic behavior mathematically. Both aspects are sides of the same coin. The quantum aspects arise from the choice of terminology in the mathematical structural analysis rather than from innately physical environments. Yet, much structural insight is gained from the study of classical interaction and decision systems in terms of  quantum theoretical language.

The key in the analysis is the symmetry decomposition of the matrix of inter\-action coefficients $\alpha_{xy}$ associated with a state $\alpha$ of an interaction system. The decomposition allows us to represent states by self-adjoint matrices. Spectral decomposition then shows that interaction states have real eigenvalues. Moreover, each state is represented by a linear combination of pure density matrices. The connection to physical quantum models is now immediate: The axioms of quantum theory stipulate that quantum states can be described as convex combinations of pure density matrices. It follows that interaction systems subsume (finite-dimensional) quantum systems as special cases. We develop the mathematical details in Section~\ref{sec:interaction-systems}.  This section treats also measurements on interaction systems as linear functionals on the space of interaction matrices and provides a probabilistic interpretation that is commensurate with the probabilistic interpretation of measurements in standard physical quantum theory.

Section~\ref{sec.decision-analysis} approaches a binary decision system with a single decision maker as a system with two ''interacting'' alternatives. Decision systems with $n$ decision makers arise as tensor products of $n$ single decision systems. Again, the relation with quantum models is immediate: The states of an $n$-decision system are described by complex $n$-dimensional vectors of unit length in exactly the same way as such vectors are assumed to describe the states of an $n$-dimensional quantum system in the Schr\"odinger picture of quantum theory.

The latter yields a very intuitive interpretation of quantum games: The players select their strategies as to move some decision makers into accepting an alternative that offers financial rewards to the individual players, where the financial reward is given as a linear functional on the space to decision states (Section~\ref{sec:decision-games}).

The choice of a set-theoretic representation of the decision states shows how classical cooperative TU games may be viewed as particular decision states of a finite set of decision makers. The associated probabilistic interpretation exhibits in particular Aubin's~\cite{Aubin81} fuzzy games as decision instances where the decision makers (or ''players'') take their decisions probabilistically but independently from one another.

The model of quantum computing\footnote{see, \emph{e.g.}, Nielsen and Chang~\cite{nielsen-chuang00}} views computation as the application of certain linear operators on quantum states. We point out in Section~\ref{sec:linear-transformations} how well-known linear transformations (\emph{e.g.}, M\"obius or Walsh transform) in cooperative games theory arise as tensor products of simple transformations on single decision systems. Moreover, we show that the classical Fourier transform is well-defined in the context of interaction systems, while it does not seem to exist when attention is restricted to the classical space of TU games.

We finally present a linear model for the evolution of interaction systems and discuss its ''Markovian'' aspects. In addition to the application examples in the course of the treatment, the appendix contains a worked out example of an interaction system with two agents.

\section{Interaction systems}\label{sec:interaction-systems}
An \emph{interaction system} is a pair $\mathfrak X =(X,\cal A)$, where $X$ is a set of \emph{agents} and $\cal A$ a set of \emph{states} such that for any state $\alpha\in \cA$ and  agents $x,y\in X$, there is a well-defined \emph{interaction coefficient} $\alpha_{xy}\in \R$.  The corresponding matrix $A=[\alpha_{xy}]\in \R^{X\times X}$ of {interaction coefficients}  reflects the \emph{interaction instance} relative to the state $\alpha$. Interpreting the diagonal elements $A_{xx}$ (interactions of the agents $x$ with themselves) as the activity levels of the individual agents in state $\alpha$, we will refer to $A$ also as an \emph{activity matrix}.

$A^T$ denotes the transpose of $A$. We do not necessarily assume interaction symmetry $A_{xy}=A_{yx}$ (\emph{i.e.}, $A^T \neq A$ may hold). Moreover, we allow ``interaction'' and ``agent''  to be taken in the most general sense. So agents can be physical or more abstract entities. However, we assume finiteness
\begin{itemize}
\item $|X|<\infty$
\end{itemize}
throughout this investigation. We give some examples of interaction systems.

\begin{example}[Buyers and sellers]\label{ex:buse}
The set $X$ of agents is partitioned into buyers and sellers. Let $t_{xy}$ be the amount of transaction done between $x$ and $y$. So $t_{xy} = 0$ holds if $x$ and $y$ are both buyers or both are
sellers. In general, we have $t_{xy} + t_{xy} =0$. It follows that the corresponding transaction state is described by the transaction matrix $A =[t_{xy}]$. $A$ is skew-symmetric (\emph{i.e.}, $A^T = -A$ holds).
\end{example}

\begin{example}[Communication networks]\label{ex:cone}
Assume that agents in $X$ can communicate pairwise and model this ability by a (directed or undirected) graph or, equivalently, by a matrix $A$ which represents the current state of communication, where $A_{xy}$ indicates the level (quality) of communication (or information) flowing from $x$ to $y$. Note that $A$ need not be symmetric.
\end{example}

\begin{example}[Influence networks]\label{ex:inne}
Assume that the set $X$ of agents forms a network and that agents communicate or
interact through this network. Through communication or interaction, the opinion
of the agents on a given topic or decision to be made may be subject to change due to mutual
influence\footnote{see, {\it e.g.}, Grabisch and Rusinowska \cite{grru1,grru2}}. The corresponding
influence matrix $A$ reflects the amount and type of influence that
agents exert among themselves. More precisely, a nonzero coefficient $A_{xy}$ indicates
that agent $x$ listens or is sensitive to the opinion of agent $y$. $A_{xy}>0$
measures the extent to which $x$ follows the opinion of $y$ (with $A_{xy}=1$
meaning that $x$ fully follows $y$) and $A_{xy}<0$ indicates the extent to which
$x$ will adopt an opinion opposite to the opinion of $y$. $A$ need not be symmetric nor skew-symmetric.
\end{example}

\begin{example}[Cooperative games]\label{ex:coga}
Let $N$ be the set of agents (or ''players''), involved in a cooperative effort
in view of achieving some goal. For any \emph{coalition} $S\subseteq N$ of players,
$v(S)$ is the quantitative result of the cooperation of the members of $S$
(assuming that the players in $N\setminus S$ show no activity): achieved
benefit, saved cost, {\it etc.} Game theory models such a situation by a
function $v:2^N\rightarrow \R$ (TU-game). In our setting, this situation can be expressed by letting $X=2^N$ be the collection of all possible coalitions and associating with $v$ the diagonal activity matrix $V$ with diagonal coefficients $V_{SS}=v(S)$ (and $V_{ST}=0$ for
$S\neq T$).

We revisit cooperative games from a different point of view (namely as binary decision systems with $|N|$ decision makers)  in Section~\ref{sec:n2} below.
\end{example}

\begin{example}[Interaction in $2$-additive games]\label{ex:in2a}
Take $X$ to be the set of players of a TU-game $v:2^X\rightarrow
\R$. Grabisch and Roubens \cite{grro99} introduced the notion of an \emph{interaction
  index} to model the interaction inside any coalition $S\subseteq X$. $v$ is said to be
  \emph{$k$-additive} if the interaction in coalitions of size greater than $k$ is zero.
  It follows that $2$-additive games are completely determined by the interaction index $I_{xy}$
  for pairs of agents together with the interactions $I_x$ of singletons, which corresponds to their
  Shapley value. The resulting interaction matrix $I$, with coefficients $I_{xy}$ for any $x\neq
y$ and $I_{xx}=I_x$ for $x\in X$, is symmetric. The index $I_{xy}$ was initially proposed by
Owen~\cite{owe72} under the name ``co-value''.

\end{example}
\begin{example}[Bicooperative games]\label{ex:biga}
The situation of Example~\ref{ex:coga} can be refined  by allowing that a cooperating coalition $S$, another coalition $T\subseteq N\setminus S$ to exist that opposes the
realization of the goal of $S$ (with the players in $N\setminus (S\cup T)$ being
inactive). Such a situation gives rise to a so-called \emph{bicooperative games}\footnote{see
Bilbao {\it et al.} \cite{bifejile00}, Labreuche and Grabisch \cite{lagr08}} and is
usually modelled by a \emph{biset function} $v$ that assigns to any pair $(S,T)$
of disjoint coalitions a quantity $v(S,T)$. With $X=2^N$, $v$  would be represented in our setting by a (non-diagonal) matrix $V$ having coefficients $V_{ST}=v(S,T)$ for any disjoint $S,T\in 2^N$, and $V_{ST}=0$ otherwise. Note that $V$ is neither symmetric nor skew-symmetric in
general.
\end{example}

Returning to a general interaction system $\mathfrak X =(X,\mathcal A)$,  recall that the space $\R^{X\times X}$ of all possible interaction matrices is a $|X|^2$-dimensional euclidian vector space. Moreover, $\R^{X\times X}$ is a real Hilbert space relative to the inner product
$$
(A|B\) = \sum_{x,y\in X} A_{xy}B_{xy} = \sum_{x\in X} (A^TB)_{xx} = \tr(A^TB),
$$
where $\tr(C)$ denotes the trace of a matrix $C$. The associated norm is the so-called \emph{Frobenius norm}
$$
  \|A\| = \sqrt{\(A|A\)} =\sqrt{\sum_{x,y\in X} |A_{xy}|^2}.
$$
We define $\|\alpha\| = \|[\alpha_{xy}]\|$ as the \emph{norm} of the state $\alpha\in\cA$.

\subsection{Symmetry decomposition and hermitian representation}\label{sec:symmetry-decomposition}
Denote by $\S_X^+$ the subspace of symmetric and by $\S_X^-$ the subspace of skew-symmetric matrices in $\R^{X\times X}$. For any $C\in \S_X^+$ and $B\in \S_X^-$,  one has
$$
\(C|B\) = \(C^T|B^T\) = -\(C|B\) \quad\mbox{and hence}\quad \(C|B\) = 0.
$$
Setting  $A^+ = \frac12(A+A^T)\in\S_X^+$ and  $A^- = \frac12(A-A^T)\in\S_X^-$, one finds that $A\in \R^{X\times X}$ is the superposition
\begin{equation}\label{eq.symmetry-decomposition}
 A = A^+ +A^-
\end{equation}
of a symmetric and a skew-symmetric matrix. It follows that $\S_X^+$ and $\S_X^-$ are orthogonal complements in $\R^{X\times X}$ and that the symmetry decomposition (\ref{eq.symmetry-decomposition}) of $A$ is uniquely determined and obeys  the pythagorean law
$$
   \|A\|^2 = \|A^+\|^2 + \|A^-\|^2.
$$

A convenient way of keeping track of the symmetry properties of $A$ is its \emph{hermitian representation} with the complex coefficients $\hat{A}_{xy} = A_{xy}^+ +\im A^-_{xy}$
\begin{equation}\label{eq.hermitian-representation}
     \hat{A} = A^+ +\im A^- \in \C^{X\times X},
\end{equation}
where $\C=\R+\im \R$ denotes the field of complex numbers with the imaginary unit $\im = \sqrt{-1}$.
So $\hat{A} = A$ holds if and only if $A$ is symmetric. Moreover, the \emph{hermitian space}
$$
   \H_X = \{A^++\im A^-\mid A\in \R^{X\times X}\} = \S_X^+ +\im \S_X^-  \subseteq \C^{X\times X}
$$
is isomorphic with $\R^{X\times X}$ as a real Hilbert space. Recalling the \emph{conjugate} of the complex matrix $C = D+\im F$ with $D,F\in \R^{X\times X}$, as the matrix $\ov{C} = D-\im F$ and the \emph{adjoint} as the matrix $C^* = \ov{C}^T$, the inner product becomes
$$
  \(A|B\) = \tr(A^TB) = \tr(\hat{A}^*\hat{B}) =\(\hat{A}|\hat{B}\) \quad(A,B\in \R^{X\times X}).
$$

\medskip
A readily verified but important observation identifies the hermitian matrices as precisely the self-adjoint matrices:
$$
    C \in \H_X \quad \Longleftrightarrow\quad C^* =C \quad(C\in \C^{X\times X}).
$$

The fact that $\R^{X\times X}$ and $\H_X$ are isomorphic Hilbert spaces allows us to view the interaction matrices $A$ and their hermitian representations $\hat{A}$ as equally valid representatives of interaction states. We denote by
$$
\hat{\alpha} = \hat{A} = A^+ +\im A^-
$$
the hermitian representation of the state $\alpha$ with interaction matrix $A=[\alpha_{xy}]$.
\subsubsection{Binary interaction}\label{sec:binary-interaction}
We illustrate the preceding concepts with the interaction of just two agents $x,y$,
\emph{i.e.}, $X=\{x,y\}$. A basis for the symmetric space $\S_X^+$ is given by
$$
   I=\begin{pmatrix} 1&0\\0&1\end{pmatrix}, \pi_1 =\begin{pmatrix} 1&0\\0&-1\end{pmatrix}, \pi_2 =\begin{pmatrix} 0&1\\1&0\end{pmatrix}.
$$
The skew-symmetric space $\S_X^-$ is $1$-dimensional and generated by
$$
\pi_3 =\begin{pmatrix} 0&-1\\1&0\end{pmatrix}.
$$
Thinking of the interaction of an agent with itself as its activity level, one can interpret these matrices as follows:
\begin{enumerate}
\item $I$: no proper interaction, the two agents have the same unit activity level.
\item $\pi_1$: no proper interaction, opposite unit activity levels.
\item $\pi_2$: no proper activity, symmetric interaction: there is a unit ''interaction flow'' from $x$ to $y$ and a unit flow from $y$ to $x$.
\item $\pi_3$: no proper activity, skew-symmetric interaction: there is just a unit flow from $x$ to $y$ or, equivalently, a $(-1)$-flow from $y$ to $x$.
\end{enumerate}
The corresponding hermitian representations are $\hat{I} = I$, $\hat{\pi}_1 = \pi_1$, $\hat{\pi}_2 = \pi_2$ and
$$
   \hat{\pi}_3 = \im\pi_3 = \begin{pmatrix} 0&-\im\\ \im&0\end{pmatrix}.
$$

\begin{remark}\label{r.Pauli}  The self-adjoint matrices $\hat{\pi}_1,\hat{\pi}_2,\hat{\pi}_3$ are the well-known \emph{Pauli spin matrices} that describe the interaction of a particle with an external electromagnetic field in quantum mechanics. Details of the underlying physical model, however, are irrelevant for our present approach. Our interaction analysis is purely mathematical. It applies to economic and game theoretic contexts equally well.
\end{remark}

\medskip\label{r.complex-numbers}
\begin{remark} The relation $\pi_3^2 = -I$ (\emph{i.e.}, ''$\pi_3 = \sqrt{-I}$'') exhibits $\pi_3$ in the role of an ''imaginary unit'' in $\R^{2\times 2}$. Indeed, the $2$-dimensional matrix space
$$
  \cC = \{aI +b\pi_3\mid a,b\in \R\} \subseteq \R^{2\times 2}
$$
is closed under matrix multiplication and algebraically isomorphic with the field $\C$ of complex numbers.
\end{remark}

\subsection{Spectral theory}\label{sec:spectral-theory}
It is a well-known fact in linear algebra that a complex matrix $C\in \C^{X\times X}$ is self-adjoint if and only if there is a diagonal matrix $\Lambda$ with diagonal elements $\lambda_x\in \R$ and a \emph{unitary} matrix $U\in \C^{X\times }$ (\emph{i.e.}, $U^* = U^{-1}$) that \emph{diagonalizes} $C$ in the sense
\begin{equation}\label{eq.diagonalization}
      \Lambda = U^*CU .
\end{equation}
The real numbers $\lambda_x$ are the eigenvalues and form the \emph{spectrum} of $C\in \H_X$. The corresponding column vectors $U_x$ of $U$ are eigenvectors of $C$ and yield a basis for the complex vector space $\C^X$.

\medskip
If $\alpha$ is a state of $\mathfrak X=(X,\mathcal A)$, we refer to the eigenvalues of its self-adjoint representation $\hat{\alpha}$
simply as the \emph{eigenvalues of $\alpha$}. A state thus has always real eigenvalues.
If the interaction matrix $[\alpha_{xy}]$ is symmetric, then $\hat{\alpha} = [\alpha_{xy}]$ holds and the eigenvalues of $[\alpha_{xy}]$ and $\hat{\alpha}$ coincide. In general, however, an interaction matrix $[\alpha_{xy}]$ does not necessarily have real eigenvalues.

\medskip
The diagonalization (\ref{eq.diagonalization}) implies the \emph{spectral representation}
\begin{equation}\label{eq.spectral-representation}
       C = \sum_{x\in X} \lambda_x U_xU^*_x
\end{equation}
with pairwise orthogonal vectors $U_x\in \C^X$ of unit length $\|U_x\|= \sqrt{U^*_xU_x} = 1$ and real numbers $\lambda_x$. (\ref{eq.spectral-representation}) shows that the members of $\H_X$ are precisely the real linear combinations of (self-adjoint) matrices of the form
$$
    uu^* \quad\mbox{with}\quad u\in \C^X \quad\mbox{s.t.}\quad \|u\|^2 =u^*u = 1.
$$

\begin{remark} In quantum theory, a matrix of the form $uu^*$ with $u\in \C^X$ of length $\|u\|= 1$ is termed a \emph{pure density (matrix)}. In the so-called \emph{Heisenberg picture}, the \emph{states} of a $|X|$-dimensional quantum system are thought to be represented by the convex linear combinations of pure densities. We thus find that the states of a general
  interaction system are represented by linear (but not necessarily convex) combinations of pure densities.
\end{remark}

\subsection{Measurements}\label{sec:measurements}
By a \emph{(linear) measurement} on the interaction system $\mathfrak X =(X,\mathcal{A})$, we understand
a linear functional $f:\R^{X\times X}\to \R$ on the space of all possible interaction instances. $f(A)$ is the value observed when the agents in $X$ interact according to $A$. So there is a matrix $F\in \R^{X\times X}$ such that
\begin{equation}\label{eq:meas}
   f(A) = \(F|A\) = \(\hat{F}|\hat{A}\) \quad\mbox{for all $A\in \R^{X\times X}$.}
\end{equation}

\begin{remark}
The second equality in (\ref{eq:meas}) shows that our measurement model is compatible with the measurement model of quantum theory, where a ''measuring instrument'' relative to a quantum system is represented by a self-adjoint matrix $\hat{F}$ and produces the value $\(\hat{F}|\hat{A}\)$ when the system is in the quantum state $\hat{A}$. However, also many classical notions of game theory can be viewed from that perspective.
\end{remark}
We give some illustrating examples.

\begin{itemize}
\item {\bf Cooperative games.} A \emph{linear value} for a player in a cooperative game ({\it sensu} Ex.~\ref{ex:coga}), is a linear functional on the collection of diagonal interß-action matrices $V$.
    Clearly, any such functional extends linearly to all interaction matrices $A$. So the Shapley value \cite{sha53} (or any linear value
  (probabilistic, Banzhaf, egalitarian {\it etc.}) can be seen as a
  measurement. Indeed, taking the example of the Shapley value, for a given
  player $i\in N$, the quantity
\[
\phi_i(v) = \sum_{S\subseteq N\setminus i}\frac{(n-s-1)!s!}{n!}(v(S\cup i) -v(S))
\]
with $s=|S|$, acts linearly on the diagonal matrices $V$ representing the characteristic function $v$. Similar conclusions apply to  bicooperative games.
\item {\bf Communication Networks.} The literature on graphs and networks\footnote{
  see, {\it e.g.}, Jackson \cite{jac08}} proposes various measures for the centrality (Bonacich centrality, betweenness {\it etc.})
  or prestige (Katz prestige index) of a given node in the graph,
  taking into account its position, the number of paths going through it,
  {\it etc.} These measures are typically linear relative to the incidence matrix of the
  graph and thus represent measurements.
\end{itemize}
Further important examples arise from payoff evaluations in $n$-person games below (\emph{cf.} Remark~\ref{r.linear-payoff}).

\subsubsection{Probabilistic interpretation}\label{sec:prob-interpretation1}
If $\hat{A}= \sum_{x\in X} \lambda_x U_xU^*_x$ is the spectral representation of the self-adjoint matrix $\hat{A}$, then the measurement (\ref{eq:meas}) takes the form
\begin{equation}\label{eq.measurement}
   f(A) = \sum_{x\in X} \lambda_x \(\hat{F}|U_x U^*_x\) =   \sum_{x\in X}\sum_{y\in X}\lambda_x \mu_y\(V_yV_y^*|U_xU_x^*\),
\end{equation}
where $\hat{F}=\sum_{y\in X} \mu_y V_yV_y^*$ is the spectral representation of $\hat{F}$. Formula (\ref{eq.measurement}) has an immediate intuitive probabilistic interpretation, well-known in quantum theory, when we set
$$
    p_{xy} = \(V_yV_y^*|U_xU_x^*\) = \tr(V_yV_y^*U_xU_x^*\).
$$
Indeed, Lemma~\ref{l.measure-probabilities} (below) implies $p_{xy} \geq 0$. Moreover, since the $V_y$ and $U_x$ yield unitary bases of $\C^X$, we have
$$
\sum_{x,y\in X}p_{xy} = \sum_{x\in X}\sum_{y\in X} \(V_yV_y^*|U_xU_x^*\) =\sum_{x\in X} \(I|U_xU_x^*\) = \(I|I\) = 1.
$$
\emph{i.e.}, the $p_{xy}$ constitute a probability distribution on the set $X\times X$ of pairs of agents $(x,y)$ and
$$
f(A)=\sum_{x,y\in X} \lambda_x\mu_y p_{xy}
$$
is the expected value of the corresponding eigenvalue products $\lambda_x\mu_y$.

\medskip
\begin{lemma}\label{l.measure-probabilities} Let $u,v\in \C^k$ be arbitrary vectors with complex coordinates $u_i$ and $v_j$. Then $\(V|U\) \geq 0$ holds for the ($k\times k$)-matrices $U=uu^*$ and $V=vv^*$.
\end{lemma}

\Pf Let $z=\sum_{j} \ov{u}_jv_j$. Since $V_{ij} = v_i\ov{v}_j$ and $U_{ij} =  u_i\ov{u}_j$, one finds
$$
\(V|U\) = \sum_{i=1}^k\sum_{j=1}^k \ov{v_i}v_j u_i\ov{u}_j = \sum_{i=1}^k \ov{v}_iu_i \cdot z = \ov{z}z = |z|^2 \geq 0.
$$
\qed

\section{Decision analysis}\label{sec.decision-analysis}
Consider $n$ decision makers, each of them facing a choice between two alternatives. Depending on the context, the alternatives can have different interpretations: 'no' or 'yes' in a voting context, for example, or 'inactive' or 'active' if the decision maker is a player in a game theoretic or economic context {\it etc.}. Moreover, the alternatives do not have to be the same for all decision makers. Nevertheless, it is convenient to denote the alternatives simply by the symbols '$0$' and '$1$'. This results in no loss of generality.

\subsection{The case $n=1$}\label{sec:sinage}
We start with a single decision maker with the binary alternatives '$0$' and '$1$'. Let $d_{jk}\in \R$ be a measure for the tendency of the decision maker to consider the choice of alternative $j$ but to possibly switch to alternative $k$. The $4$-dimensional coefficient vector  $d=(d_{00},d_{10},d_{01},d_{11})$ represents this {state} of the decision maker. Moreover, in view of the isomorphism
$$
   \R^4 \;\cong\; (\R+\im \R) \times (\R+\im \R),
$$
we can represent $d$ by a pair of complex numbers $\delta_j$:
$$
    \hat{d} = (d_{00}+ \im d_{10} , d_{01}+ \im d_{11}) =(\delta_0,\delta_1).
$$
In the non-trivial case $d\neq 0$, there is no loss of generality when we assume $d$ (and hence $\hat{d}$) to be unit normalized
$$
   \|d\| = \sqrt{d_{00}^2+d_{10}^2 + d_{01}^2 +d_{11}^2} = \|\hat{d}\| = 1.
$$
So we think of the unit vector set
$$
\cS = \{\delta =(\delta_0,\delta_1)^T \in \C^2\mid \|\delta\|^2 = |\delta_0|^2 +|\delta_1|^2 =1\}
$$
as the set of \emph{(proper) states} of the decision maker.

\subsubsection{Decisions and interactions}\label{sec:decisions-interactions}
While a vector state $\delta=(\delta_0,\delta_1)^T\in \cS$ describes a state of the decision maker, the decision framework implies a natural interaction system in the set $X=\{0,1\}$ of alternatives. Indeed, the matrix
\begin{equation}\label{eq.decision-density}
   D = \delta \delta^* = \begin{pmatrix} \delta_0\ov{\delta}_0 & \delta_0\ov{\delta}_1 \\ \delta_1\ov{\delta}_0 &\delta_1\ov{\delta}_1\end{pmatrix} =
   D = \delta \delta^* = \begin{pmatrix} |\delta_0|^2 & \delta_0\ov{\delta}_1 \\ \delta_1\ov{\delta}_0 &|\delta_1|^2\end{pmatrix}
\end{equation}
associates with the decision state $\delta$ a ''quantum state'' with density $D$, which may  be interpreted as the self-adjoint representation of an interaction state on $X$. The latter exhibits the alternatives '0' and '1' as interacting agents in their own right in the decision process.

\subsubsection{Decision probabilities}\label{sec:decision-probabilities} The decision state vector  $\hat{d} = (d_{00}+ \im d_{10} , d_{01}+ \im d_{11}) =(\delta_0,\delta_1)\in \cS$  defines a pro\-bability distribution $p$ on $X^2 =\{0,1\}^2$ with probabilities $p^{(d)}_{jk} = d_{jk}^2$. Accordingly, the probability for the alternative '$k$' to be accepted is
$$
   p_k^{(d)} = p^{(d)}_{0k} + p^{(d)}_{1k} =  |\delta_k|^2 \quad(k=0,1).
$$
We arrive at the same probabilities when we apply suitable measurements in the sense of Section~\ref{sec:measurements} on the interaction system on $X$ with the self-adjoint state representation $D$ as in (\ref{eq.decision-density}). To see this, let us measure the activity level of $k$ (\emph{i.e.}, the interaction of $k$ with itself).
This measurement corresponds to the application of the matrix $F_k = e_ke_k^T$, where $e_k$ is the $k$th unit vector in $\R^2$. The expected value of this measurement is
$$
\(F_k|D\) = |\delta_k|^2 =  p_k^{(d)} .
$$
Let us now take another look at  earlier examples:
\begin{enumerate}
\item Influence networks: While it may be unusual to speak of ''influence'' if there is only  a single agent, the possible decisions of this agent are its opinions ('yes' or 'no'),
  and the state of the agent hesitating between 'yes' and 'no' is described by
  $(\delta_0,\delta_1)\in\C^2$ (with probability $|\delta_0|^2$ to say 'no' and
  $|\delta_1|^2$ to say 'yes').
\item Cooperative games: As before, ''cooperation'' may sound awkward in the case of a
  single player. However, the possible decisions of the player are relevant and amount to being either active or inactive in the given game. Here,  $(\delta_0,\delta_1)$ represents a
  state of deliberation of the player that results in the probability $p_k^{(d)}$ for being active (resp. inactive).
\end{enumerate}

\subsubsection{Quantum bits}\label{sec:qbits}
Denote by $|j\)$ the event that the decision maker selects alternative $j$. Thinking of $|0\)$ and $|1\)$ as independent generators of a $2$-dimensional complex vector space, we can represent the decision states equally well as formal linear combinations
\begin{equation}\label{eq.q-bit}
   \delta = \delta_0|0\) + \delta_1|1\) \quad ( (\delta_0,\delta_1)^T\in \cS).
\end{equation}
The expression (\ref{eq.q-bit}) is the representation of the states of a $1$-dimensional quantum system in the so-called \emph{Schr\"odinger picture} of quantum mechanics. Strong arguments have been put forward\footnote{see, \emph{e.g.},
Penrose~\cite{penros94}} that the decision process of the human mind should be
considered to be a quantum process and a binary decision state should thus be described as a quantum state.

In the theory of quantum computing\footnote{see, \emph{e.g.}, Nielsen and Chuang
  \cite{nielsen-chuang00}}, an expression of the form (\ref{eq.q-bit})
represents a \emph{quantum bit}. $|0\)$ and $|1\)$ correspond to the boolean
bits of classical computation.

\subsubsection{Non-binary alternatives}\label{sec:non-binary-alternatives1}
It is straightforward to generalize the decision model to $k>1$ alternatives '$0$', '$1$', $\ldots$, '$k-1$'. We let $|j\)$ denote the event that '$j$' is selected. Interpreting '$j$' as representing an ''activity'' at the $j$th of $k$ possible levels, the decision maker is essentially a player in a \emph{multichoice game} in the sense of Hsiao and Raghavan~\cite{hsra92}. Decision states are now described by formal linear combinations
$$
\delta = \sum_{j=0}^{k-1} \delta_j|j\) \quad\mbox{with $\delta_j\in \C$ s.t. $ \|\delta\|^2 =\D\sum_{j=0}^{k-1} |\delta_j|^2 = 1$}
$$
and $|\delta_j|^2$ being the  probability for alternative $j$ to be taken.

\subsection{The case $n\geq 2$}\label{sec:n2}
We now extend our framework to the case of several decision makers and consider first the binary case. We
let $\mathfrak D$ denote of the decision system for $n=1$ that
can be in any state as in Section~\ref{sec:sinage}. The system of $n >1$
decision makers is then given by the $n$-fold tensor product:
$$
  \mathfrak D^n = \mathfrak D \otimes \cdots \otimes \mathfrak D \quad(\mbox{$n$ times}).
$$
To make this precise, recall the tensor product of a vector space $V$ with a
  fixed basis $B=\{b_1,\ldots,b_n\}$ and a vector space $W$ with basis $G
  =\{g_1,\ldots,g_m\}$ over the common real or complex
  scalar field $\F$. We first define the set $B\otimes G$ of all (formal)
  products $b_i\otimes g_j$ and extend the products linearly to pro\-ducts of
  vectors in $V$ and $W$, \emph{i.e.}, if $v =\sum v_i b_i$ and $w =\sum w_j
  g_j$, then
\begin{equation}\label{eq.tensor-product}
v\otimes w =\big(\sum_{i=1}^n v_ib_i\big)\otimes \big(\sum_{j=1}^m w_j g_j\big) = \sum_{i=1}^n\sum_{j=1}^m (v_iw_j) b_i\otimes g_j.
\end{equation}
So  a coordinate representation of $v\otimes w$ is given by the ($n\times m$)-matrix with coefficients $v_iw_j$. Moreover, we note the multiplication rule for the norm:
$$
  \|v\otimes w\|^2 = \|v\|^2 \|w\|^2.
$$
The tensor product $V\otimes W$ is the $nm$-dimensional vector space with basis $B\otimes G$.
It is well-known that the tensor product of vectors is associative and distributive and linear relative to the scalar field $\F$.

We define $\mathfrak D^n$ as the system with the state set
$$
     \cS^n = \{\delta \in \C^2\otimes \cdots \otimes \C^2\mid \|\delta\| = 1\}.
$$
The construction becomes immediately clear when bit notation is employed. For any sequence $j_1\ldots j_n\in \{0,1\}^n$, we define the \emph{$n$-bit}
\begin{equation}\label{eq.n-bit}
   |j_1\ldots j_n\) = |j_1\)\otimes \cdots\otimes |j_n\).
\end{equation}
The states of $\mathfrak D^n$ now take the form of linear combinations
\begin{equation}\label{eq.n-decision-state}
    \delta = \sum_{k\in \{0,1\}^n} \delta_k |k\) \quad\mbox{with $\delta_k\in \C$ and $\sum_k |\delta_k|^2 = 1$.}
\end{equation}
In state $\delta$, the (joint) decision $k\in \{0,1\}^n$ is reached with probability $|\delta_k|^2$.

\begin{remark}
Note the relation with the model of $k$ decision alternatives in Section~\ref{sec:non-binary-alternatives1}: The representation of $\delta$ in (\ref{eq.n-decision-state}) allows us to view $\mathfrak D^n$ as the context of a single decision maker that faces the $2^n$ alternatives $|j\)$.
\end{remark}

As an illustration, consider the case $n=2$. The decision system $\mathfrak D^2=\mathfrak D\otimes \mathfrak D$ relative to a pair of binary alternatives represents the states in the form
$$
  \delta = \delta_{00}|00\) + \delta_{10}|10\) + \delta_{11}|11\) +\delta_{01}|01\)\quad(\delta_{ij}\in \C, \sum_{j,k\in \{0,1\}} |\delta_{jk}|^2 = 1).
$$
In particular, we have
$$
(\alpha_0 |0\) + \alpha_1|1\))\otimes(\beta_0 |0\) + \beta_1|1\)) =\sum_{i,j\in \{0,1\}} \alpha_i\beta_j |ij\).
$$

In general, it is often convenient to represent states in set theoretic notation. So we let $N=\{1,\ldots, n\}$ and identify $n$-bits with subsets:
$$
  k = j_1\ldots j_n \in \{0,1\}^n\quad \longleftrightarrow\quad K =\{\ell\in N\mid i_\ell = 1\},
$$
which yields the state representation
\begin{equation}\label{eq.set-representation}
\delta = \sum_{K\subseteq N} \delta_K |K\) \quad(\delta_K\in \C, \sum_{K\subseteq N} |\delta_K|^2 = 1).
\end{equation}
The latter allows us to view a state simply as a valuation $\alpha:2^N\to \C$ that assigns to each subset $K\subseteq N$ a complex number $\alpha_K\in \C$. This generalizes the classical model of cooperative games with $n$ players to complex-valued cooperative games in a natural way.

The probabilistic interpretation says that the decision $n$-tuple $k=i_1\ldots i_n$ is realized in a non-trivial state $\alpha$ of $\mathfrak D_n$ with probability
\begin{equation}\label{eq:proba}
p^\alpha_k = \frac{|\alpha_k|^2}{\|\alpha\|^2}\;, \quad\mbox{where}\quad \|\alpha\|^2 = \sum_{k\in \{0,1\}^n} |\alpha_k|^2.
\end{equation}
Coming back to our examples, a similar interpretation as for the single agent
case can be given. Specifically, in the case of the influence network in state
$\alpha$, $p^\alpha_k$ in (\ref{eq:proba}) yields the probability that a given
coalition $K$ of agents says 'yes' and the other agents say 'no', while in the
case of cooperative games, $p^\alpha_k$ is the probability for a
given coalition $K$ to be active while the remaining players are not.

\subsubsection{Entanglement and fuzzy systems}\label{sec:entanglement}
We say that a state $\delta$ of $\mathfrak D^n$ is \emph{reducible} if there is a $0 < m<n$ such that
$$
   \alpha = \alpha\otimes \beta\quad\mbox{for some state $\alpha$ of $\mathfrak D^m$ and state $\beta$ of $\mathfrak D^{n-m}$}.
$$
In this reducible case, the state $\delta$ arises from lower dimensional states $\alpha$ and $\beta$ that are probabilistically independent. Indeed, we have $\|\delta\| = \|\alpha\|\cdot\|\beta\|$, $\delta_{k\ell} = \alpha_k\beta_\ell$ and hence
$$
   p^\delta_{k\ell} = p^\alpha_k\cdot p^\beta_\ell\quad\mbox{for all $k\in \{0,1\}^m, \ell\in\{0,1\}^{n-m}$.}
$$
Already for $n=2$, it is easy to see that $\mathfrak D^n$ admits irreducible states. Such states are said to be \emph{entangled}.

Aubin~\cite{Aubin81} has introduced the model of \emph{fuzzy cooperation} of a set $N$ of $n$ players as a system whose states are characterized by real vectors
$$
     w = (w_1,\ldots,w_n) \quad\mbox{with}\quad 0\leq w_j\leq 1.
$$
In state $w$, player $j$  decides to be active in the cooperation effort with probability $w_j$ and inactive to be with probability $1-w_j$.  A coalition $S\subseteq N$ is assumed to be formed with probability
$$
    w(S) = \prod_{s\in S}\prod_{t\in N\setminus S} w_i(1-w_t)
$$
(see also Marichal~\cite{marichal09}). So the players act independently in each ''fuzzy'' state $w$. Entangled states are thus excluded. In particular, one sees that our model of interactive decisions is strictly more general than the fuzzy model.

\subsection{Linear transformations}\label{sec:linear-transformations}
Let again $B=\{b_1,\ldots,b_n\}$ be a basis of the vector space $V$ and $G =\{g_1,\ldots,g_m\}$ a basis of the vector space $W$ over the common scalar field $\F$ and consider linear operators $L:V\to V$ and $M:W\to W$. The assignment
$$
L\otimes M(b\otimes g) = (Lb)\otimes (Mg) \quad((b,g)\in B\times G)
$$
extends to a unique linear operator $L\otimes M: V\otimes W\to V\otimes W$ since the elements $b\otimes g$ form a basis of $V\otimes W$. The operator $L\otimes M$ is the \emph{tensor product} of the operators $L$ and $M$.

For illustration, let us apply this construction to various linear operators on $\mathfrak D$ and derive several linear transformation of well-known importance in game theory.

\subsubsection{M\"obius transform}\label{sec:Moebius-transform}
The \emph{M\"obius operator} $Z$ on the state vector space of $\mathfrak D$ is the linear operator such that
$$
Z|0\) = |0\)+|1\)  \quad\mbox{and}\quad Z|1\) = |1\).
$$
$Z$ admits a linear inverse $M = Z^{-1}$, which is determined by
$$
M |1\) = |1\)  \quad\mbox{and}\quad M|1\) = |1\) -|0\).
$$
The $n$-fold tensor product $Z^n =Z\otimes \cdots \otimes Z$ acts on the states of $\mathfrak D^n$ in the following way:
$$
Z^n|i_1\ldots i_n\) = Z|i_1\)\otimes \cdots \otimes Z|i_n\).
$$
Setting $N=\{1,\ldots,n\}$ and $S = \{j\in N\mid i_j = 1\}$, the set theoretic notation thus yields
$$
Z^n|S\) = \sum_{T\supseteq S} |T\)= \sum_{T\subseteq N} \zeta_{ST}|T\)
$$
with the \emph{incidence coefficients}
$$
\zeta_{ST} = \left\{\begin{matrix} 1 &\mbox{if $S\subseteq T$,}\\ 0 &\mbox{otherwise.}\end{matrix}\right.
$$
For the inverse $M^n = M\otimes \cdots \otimes M$ of $Z^n$, one has
$$
M^n|i_1\ldots i_n\) = M|i_1\)\otimes \cdots \otimes M|i_n\)
$$
and thus for any $S\subseteq N$,
$$
M^n|S\) =  \sum_{T\supseteq S} (-1)^{|T\setminus S|}|T\) = \sum_{T\subseteq N} \mu_{ST}|T\)
$$
with the \emph{M\"obius coefficients}\footnote{see Rota~\cite{rot64} for generalizations}
$$
\mu_{ST} = \left\{\begin{matrix} (-1)^{|T\setminus S|} &\mbox{if $S\subseteq T$,}\\ 0 &\mbox{otherwise.}\end{matrix}\right.
$$

The M\"obius operator $Z^n$  has well-known game theoretic implications. With the characteristic function  $v:2^N\to\R$ of a cooperative game on $N$, we associate the state vector
$$
\tilde{v} = \sum_{S\subseteq N} v(S)|S\).
$$
$v$ is said to be a \emph{unanimity game} if $\tilde{v} = Z^n|S\)$ holds for some $S\subseteq N$. In general, we may compute the M\"obius transform
$$
 \hat{v} = Z^n\tilde{v} =\sum_{S\subseteq N} v(S)Z^n|S\) =\sum_{S\subseteq N} \hat{v}(S)|S\)
$$
which expresses the M\"obius transform $\hat{v}$ of $\tilde{v}$ as a linear combination of state vectors of unanimity games. Similarly, an application of the inverse $M^n$ yields
$$
\tilde{v} = M^n \hat{v} = \sum_{S\subseteq N} \hat{v}(S)M^n|S\) =\sum_{S\subseteq N}\sum_{T\subseteq S} (-1)^{|S\setminus T|}\hat{v}(T)|S\),
$$
which yields the \emph{M\"obius inversion formula}
$$
     v(S) = \sum_{T\subseteq S} (-1)^{|S\setminus T|}\hat{v}(T) \quad(S\subseteq N)
$$
The parameters $h^v_{ST} =  (-1)^{|S\setminus T|}\hat{v}(T)$ are known as the \emph{Harsanyi coefficients} of $v$ in cooperative game theory.

\subsubsection{Hadamard transform}\label{sec:Hadamard-transform}
The \emph{Hadamard transform} $H$ of $\mathfrak D$ is the linear transformation with the property
$$
   H|0\) = \frac{|0\) + |1\)}{\sqrt{2}} \quad\mbox{and}\quad H|1\) = \frac{|0\) - |1\)}{\sqrt{2}} .
$$
The normalizing factor $\sqrt{2}$ is chosen as to make $H$ norm preserving. Note that in both states $H|0\)$ and $H|1\)$ the alternatives $0$ and $1$ are attained with equal probability
$$
    p^{H|j\)}_0 = \frac12 =  p^{H|j\)}_1 \quad(j=0,1)).
$$
It is easy to check that $H$ is self-inverse (\emph{i.e.}, $H^{-1} = H$). If follows that also the $n$-fold tensor product $H^n = H\otimes \cdots \otimes H$ is norm preserving and self-inverse on the vector state space of $\mathfrak D^n$. In particular, the $2^n$ state vectors
$$
H^n|i_1\ldots i_n\) = H|i_1\) \otimes \cdots \otimes  H|i_n\) \quad( i_j\in \{0,1\})
$$
are linearly independent. In the set theoretic notation, one finds
$$
H^n|S\) =\frac{1}{2^{n/2}} \sum_{T\subseteq N} (-1)^{|S\cap T|} |T\).
$$
Note that in each of the $2^n$ states $H^n|S\)$ of $\mathfrak D^n$, the $2^n$ boolean states $|T\)$ of $\mathfrak D^n$ are equi-probable.

\begin{remark} Both the M\"obius as well as the Hadamard transform map state vectors of $\mathfrak D^n$ with real coefficients onto state vectors with real coefficients. In particular, these transforms imply linear transformations on the space of characteristic functions of classical cooperative TU-games.

In the classical context, $H^n$ is closely related to the
so-called \emph{Walsh transform} (\emph{cf.} Walsh~\cite{wal23}), which is also
known as \emph{Fourier transform} in the theory so so-called
\emph{pseudo-boolean functions}, \emph{i.e.}, essentially (possibly non-zero
normalized) characteristic functions\footnote{Hammer and Rudeanu \cite{haru68}}. This transform is an important tool in discrete mathematics and game theoretic analysis\footnote{see, \emph{e.g.}, Kalai~\cite{kalai02}, O'Donnell~\cite{odonnell14}}.

We stress, however, that the Hadamard transform is \emph{not} the same as the classical \emph{(discrete) Fourier transform} (below) on $\mathfrak D^n$ if $n\geq 2$.
\end{remark}

Lastly, we point out an interesting connection of the Hadamard transform with
interaction transforms of Grabisch and Roubens \cite{grro99}, already
alluded to in Example~\ref{ex:in2a}. These have been proposed in the context of
cooperative games, and are essentially of two types: the Shapley interaction
transform, which extends the Shapley value, and the Banzhaf interaction
transform, which extends the Banzhaf value. The latter is expressed as
follows.

The \emph{Banzhaf transform} of the TU-game  $v:2^N\rightarrow \R$ is the
pseudo-boolean function $I_{\rm B}$ with the values
\[
I_{\rm B}^v(S) = \Big(\frac{1}{2}\Big)^{n-s}\sum_{T\subseteq N}(-1)^{|S\setminus
  T|}v(T) \qquad (S\in 2^N)
\]
representing the quantity of interaction existing among agents in
$S$. $I_{\rm B}^v(S)>0$ indicates that the cooperation of all agents in $S$
brings more than any combination of subgroups of $S$, while  $I_{\rm B}^v(S)<0$ implies some redundancy/overlap in the cooperation of agents in $S$.

By the identitification TU-game '$v$ $\leftrightarrow$ state $\alpha$', the
Hadamard transform becomes a transform $H^v$ on TU-games with values
\[
H^v(S) =\frac{1}{2^{n/2}} \sum_{T\subseteq N} (-1)^{|S\cap T|}v(T).
\]
Now, it is easy to check that
\[
I_{\rm B}^v(S) = \frac{(-2)^{s}}{2^{n/2}}H^v(S)
\]
holds, which yields an interpretation of the Hadamard transform in terms of interaction in cooperative contexts.

\subsubsection{Fourier transformation}\label{sec:Fourier-transformation}
We briefly recall the classical discrete Fourier transform of the coordinate space $\C^k$. Setting $\omega = e^{2\pi\im/k}\in \C$, one defines the (unitary) \emph{Fourier matrix}
$$
   \Omega = \frac{1}{\sqrt{k}}\begin{pmatrix}\omega^1&\omega^2 &\ldots& \omega^{k}\\
   \omega^{2}&\omega^4 &\ldots& \omega^{2k}\\
   \vdots &\vdots &\cdots &\vdots \\ \omega^k &\omega^{2k} &\ldots&\omega^{k^2} \end{pmatrix} \in \C^{k\times k}.
$$
The \emph{Fourier transform} of $v\in \C^k$ is the vector $\Omega v\in \C^k$. Applied to decision systems, the  Fourier transform of any state vector will also be a state vector.

\begin{remark} Note that the Fourier transform of a state vector with real coefficients is not necessarily a real vector. In the language of TU games, this means that the Fourier transform of a TU game is not necessarily a TU game. Yet, the Fourier transform is well-defined and meaningful in the wider model of decision systems.
\end{remark}

The Fourier transform extends naturally to interactions. Indeed, for any matrix $M\in \C^{k\times k}$, the linear operator
$$
C \mapsto \mu(C) =  M C\ M^* \quad(C\in \C^{k\times k})
$$
preserves self-adjointness and thus acts as a linear operator on  the Hilbert space $\H_k$ of all ($k\times k$) self-adjoint matrices. In particular,  we have
$$
    C =\sum_{i=1}^k \lambda_i u_iu_i^*\;\Longrightarrow \; \mu(C) = \sum_{i=1}^k M(\lambda_i u_iu_i^*)M^* = \sum_{i=1}^k \lambda_i (Mu_i)(Mu_i)^*.
$$
Hence, if $M$ is unitary, we observe that the spectral decomposition of a self-adjoint matrix $C$ with eigenvectors $u_i$ yields a spectral decomposition of $\mu_M(C)$ with eigenvectors $Mu_i$ (and the same eigenvalues).

The choice $M = \Omega$ thus yields the Fourier transform for interaction instances, which preserves the symmetry values:
$$
    \hat{A} \mapsto \mu_\Omega(\hat{A}) = \Omega \hat{A}\Omega^* \quad(A\in \R^{k\times k}).
$$

\subsection{Decision and quantum games}\label{sec:decision-games}
An \emph{$n$-person game} $\Gamma$ involves $n$ players  $j$, each of them having a set $S_j$ of feasible strategies relative to some system $\mathfrak S$, which is assumed to be in an initial state $\alpha^{(0)}$. Each player $j$ selects a strategy $s_j\in S_j$ and the joint selection  $s=(s_1,\ldots,s_n)$ of strategies then moves the system into a final state $\alpha^{(f)}$. The reward of player $j$ is $P^{(j)}(\alpha^{(f)})$, where $P^{(j)}$ is a pre-specified real-valued functional on the state space of $\mathfrak S$.

The $n$-person game $\Gamma$ is a \emph{decision game} if it is played relative to a decision system $\mathfrak D^m$ with binary alternatives for some $m>0$. By their strategic choices, the $n$ players influence the $m$ decision makers towards a certain decision, \emph{i.e.}, the joint strategy $s$ moves the game to a decision state
$$
\delta^{(f)} = \sum_{k\in\{0,1\}^m} \delta_k^{(f)} |k\)  \quad(\delta^{(f)}_k\in\C, \sum_{k\in\{0,1\}^m} |\delta_k^{(f)}|^2 = 1).
$$
In the state $\delta^{(f)}$, the $m$ decision makers will accept the alternative $k$ with probability  $|\delta_k^{(f)}|^2$, in which case $|k\)$ is considered to be the final state of the game and player $j$ receives a pre-specified reward $p_k^{(j)}$. Hence $j$'s  \emph{expected} payoff is
\begin{equation}\label{eq.q-game-payoff}
    P^{(j)}(\delta^{(f)}) =\sum_{k\in\{0,1\}^m} p^{(j)}_k |\delta_k^{(j)}|^2.
\end{equation}

Interpreting the decision states of $\mathfrak D^m$ as the states of an $m$-dimensional quantum system and regarding $\delta^{(f)}$ as the final state of the game, we arrive at the model of a \emph{quantum game} with the payoff functionals $P^{(j)}$.

\begin{remark}\label{r.linear-payoff}
The payoff functionals $P^{(j)}$ reveal themselves as {linear measurements} if we represent decision states not as vectors $\delta$ but as density matrices $\delta\delta^*$:
$$
    P^{(j)}(\delta)  =\sum_{k\in\{0,1\}^m} p^{(j)}_k |\delta_k|^2 =\(P^{(j)}|\delta\delta^*\),
$$
where $P^{(j)}$ is now the diagonal matrix with coefficients $P_{kk}^{(j)}=   p^{(j)}_k $.
\end{remark}

In the quantum games discussed in the literature, the strategies of the individual players typically consist in the application of certain linear operations on the states\footnote{see, {\it e.g.}, Grabbe~\cite{grabbe05} or Guo {\it et al.}~\cite{guo-et-al08}}. As an illustration, consider the generalization of the classical Prisoners' Dilemma of Eisert {\it et al.}~\cite{eisert-et-al99}: :

There are $2$ players relative to the system $\mathfrak D^2 = \mathfrak D\otimes \mathfrak D$ with states given as
$$
   \delta = \delta_{00}|00\)+\delta_{01}|01\) + \delta_{10}|10\) + \delta_{11}|11\) \quad(\delta_{ij}\in \C).
$$
The game is initialized by a unitary operator $U$ and prepared into the state $\alpha^{(0)}= U|00\)$. The players select unitary operators $A$ and $B$ on $\mathfrak{D}$, whose tensor product $A\otimes B$ is applied to $\alpha^{(0)}$. A further application of $U^*$  results in the final state
$$
\alpha^{(f)} = U^*(A\otimes B)\alpha^{(0)} = U^*(A\otimes B)U|00\).
$$
The payoff coefficients $p_k^{(j)}$, with $k\in \{00,01,10,11\}$, are the coefficients of the payoff matrix in the classical Prisoners' Dilemma.

Strategic choices of operators associated with the Pauli matrices $\pi_1$ and $\pi_2$  (Section~\ref{sec:binary-interaction}), for example, would act on the states in the following way:
\begin{eqnarray*}
\pi_1 |0\) &=& |0\) \quad\mbox{and}\quad  \pi_1 |1\) = -|1\)\\
\pi_2 |0\) &=& |1\) \quad\mbox{and}\quad  \pi_2 |1\) = |0\).
\end{eqnarray*}
Eisert {\it et al.} show that the set $\{I,\pi_1,\pi_2\}$ of admissible strategies for each player can guarantee the existence of a Pareto optimal Nash equilibrium in cases where the classical variant does not admit such a Nash equilibrium.

Without going into further details, we mention that the Hadamard transform $H$ often turns out to be a powerful strategic operator in quantum games (see, \emph{e.g.}, the seminal penny flip game of Meyer~\cite{meyer99}).

\section{Markov evolutions}\label{sec:Markov-evolutions}
Interaction among agents may be time dependent. For example, opinions form over time in mutual information exchange, and game theoretic coalitions form due to changing economic conditions {\it etc.} Generally, we understand by an \emph{evolution} of an interaction system $\mathfrak X = (X,\cal A)$ in discrete time $t$ a sequence $\epsilon= (\alpha_t)_{t\geq 0}$ of states $\alpha_t\in \mathcal A$.

The evolution $\epsilon$ is \emph{bounded} if there is some $c\in \R$ such that $\|\alpha_t\|< c$ holds for all $t$. $\epsilon$ is \emph{mean ergodic} if the averages of the state representations
$$
\ov{\alpha}^{(t)} = \frac1t\sum_{m=1}^t \hat{\alpha}_m \in \H_X
$$
converge to a limit $\ov{\alpha}^{(\infty)}$. We say that $\epsilon$ is a \emph{Markov evolution} if there is a linear (''evolution'') operator $\Phi: \H_X\to\H_X$ such that
$$
 \hat{\alpha}_{t+1} = \Phi^t \hat{\alpha}_0 \;,\emph{i.e.}, \;    \hat{\alpha}_{t+1} = \Phi \hat{\alpha}_t \quad\mbox{holds for all $t\geq 0$.}
$$
A Markov evolution is mean ergodic if and only if it is bounded:

\begin{theorem}\label{t.FS} Let $\phi$ be a linear operator on $\C^k$. Then for any $a\in \C^k$, the following statements are equivalent:
\begin{enumerate}
\item[(i)] There is some $c\in \R$ such that $\|\phi^t a\|\leq c$ holds for all $t=0,1$.
\item[(ii)]The limit
$
    \ov{a}^{(\infty)} = \D\lim_{t\to\infty} \frac1t\sum_{m=1}^t a_m
$ exists.
\end{enumerate}
\end{theorem}

\Pf Theorem~2 in Faigle and Sch\"onhuth~\cite{FS}.

\qed

\medskip
The importance of these notions lies in the fact that the mean ergodic evolutions guarantee the statistical convergence of arbitrary measurements (in the sense of Section~\ref{sec:measurements}) on the evolution. More precisely, we have

\begin{corollary}\label{c.FS} Let $\phi$ be a linear operator on $\C^k$. Then for any $a\in \C^k$, the following statements are equivalent:
\begin{enumerate}
\item[(i)] The evolution sequence $(\phi^ta)_{t\geq 0}$ is mean ergodic.
\item[(ii)] For every linear functional $f:\C^k\to \C$, the statistical averages
$$
    \ov{f}^{(t)} = \frac1t\sum_{m=1}^t f(\phi^m(a))
$$
converge.
\end{enumerate}
\qed
\end{corollary}
We illustrate this model with two well-known examples.

\medskip
\paragraph{Markov chains.} Let $\pi\in\R^k$ be a probability distribution and $M\in \R^{k\times k}$ such that every ''state'' $\pi^{(t)} = M^t\pi$ is a probability distribution and hence satisfies $\|\pi^t\| \leq k$. It follows from Theorem~\ref{t.FS} that $(\pi^{(t)})_{t\geq 0}$ is a mean ergodic Markov evolution.

If $M$ is a probabilistic transition matrix (\emph{i.e.}, all columns of $M$ are probability distributions), the Markov evolution $(\pi^{(t)})_{t\geq 0}$ is a classical \emph{Markov chain}, in which case mean-ergodicity is well-known.

\medskip
\begin{remark}
In the theory of \emph{coalition formation} in cooperative game
theory, Markov chains are typically taken as underlying models for the
evolutionary formation process\footnote{see, e.g., Faigle and Grabisch \cite{fagr10}}.
\end{remark}

\medskip
\paragraph{Schr\"odinger's wave equation.} Recall that the Schr\"odinger picture of quantum mechanics describes a \emph{state evolution} of a finite-dimensional quantum system by a time dependent function $\psi$ with values $\psi(t)\in \C^k$ for some $k<\infty$, which is supposed to satisfy the differential equation
\begin{equation}\label{eq.Schroedinger}
 \frac{\partial \psi(t)}{\partial t} = \frac{\im }{\hbar}H\psi(t),
\end{equation}
with the matrix $H$ being the so-called \emph{Hamiltonian} of the system and $\hbar$  Planck's constant. Assuming $H$ to be normal  (\emph{i.e.}, $HH^* = H^*H$), one obtains the solution of (\ref{eq.Schroedinger}) in the form
$$
   \psi(t) = U_t \psi(0) \quad\mbox{(with $U_t= e^{-\im H t/\hbar}$).}
$$
Note that $U_t$ is unitary. So $\|\psi(t)\| = \|\psi(0)\|$ is bounded. If $H$ moreover is constant in time, one finds
\begin{equation}\label{eq.unitary-evolution}
  \psi(t) = U^t\psi (0)  \quad\mbox{with $U= e^{-\im H /\hbar}\in\C^{k\times k}$ unitary,}
\end{equation}
which exhibits  the (discrete) Schr\"odinger evolution $(\psi(t))_{t\geq 0}$ as mean-ergodic.

\begin{remark} More generally, one sees that arbitrary unitary evolutions in $\C^k$ are mean ergodic. This fact is well-known as \emph {von Neumann's mean ergodic theorem}.
\end{remark}

The examples show that traditional models for the evolution of interaction systems are quite restrictive and suggest to study evolutions in a wider context.

\begin{remark} For a possible infinite-dimensional evolution model that includes event observations relative to arbitrary stochastic processes as a special case, see Faigle and Gierz~\cite{FaigleGierz17}.
\end{remark}

\section*{Appendix: An example with two agents}
Let us illustrate the main notions and results introduced in the paper. For simplicity,
we consider a set $X$ two agents and interaction matrices of the form
\[
A=\begin{pmatrix} w_1 & 1-w_1 \\ 1-w_2 & w_2\end{pmatrix}
\]
with $w_1,w_2\in[0,1]$. $A$ is row-stochastic but not symmetric unless
$w_1=w_2$. $A$'s entries are generally interpreted as interaction/activity
levels, so that $w_1,w_2$ are the activity levels of agents 1 and 2, while
$1-w_1,1-w_2$ are their interaction levels. Following Example~\ref{ex:inne}, we understand
$A$ as an influence matrix. In this case, agent $i$
would listen to the opinion of the other agent with weight $1-w_i$, and to himself
with weight $w_i$.

We first apply the symmetry decomposition and find
\[
A^+=\begin{pmatrix} w_1 & 1-\frac{1}{2}W \\  1-\frac{1}{2}W & w_2\end{pmatrix},
\quad A^-=\begin{pmatrix} 0 & -\frac{1}{2}\Delta W \\ \frac{1}{2}\Delta W & 0\end{pmatrix}
\]
with the shorthand $W=w_1+w_2$ and $\Delta W=w_1-w_2$. Therefore, the
self-adjoint representation is
\[
\hat{A} = \begin{pmatrix} w_1 & 1-\frac{1}{2}W -\im \frac{1}{2}\Delta W\\
1-\frac{1}{2}W +\im \frac{1}{2}\Delta W & w_2\end{pmatrix}
\]
with the eigenvalues
\[
\lambda = \frac{W\pm \sqrt{(W-2)^2+2\Delta W^2}}{2}.
\]
The corresponding eigenvectors are
\[
u_1 = \frac{1}{\sqrt{8+2W^2-8W+4\Delta W^2-2\Delta W\sqrt{D}}}\begin{pmatrix}
  \sqrt{4+W^2-4W + \Delta W^2} \\ \displaystyle\frac{(1-\frac{1}{2}W + \im \frac{\Delta
      W}{2})(-\Delta W+\sqrt{D})}{\sqrt{1+\frac{1}{4}W^2 - W + \frac{\Delta
        W^2}{4}}}\end{pmatrix}
\]
\[
u_2 =  \frac{1}{\sqrt{8+2W^2-8W+4\Delta W^2+2\Delta W\sqrt{D}}}\begin{pmatrix}
  \sqrt{4+W^2-4W + \Delta W^2} \\ \displaystyle\frac{(1-\frac{1}{2}W + \im \frac{\Delta
      W}{2})(-\Delta W-\sqrt{D})}{\sqrt{1+\frac{1}{4}W^2 - W + \frac{\Delta
        W^2}{4}}}\end{pmatrix}
\]
One can see that calculations can become rapidly quite complex. In order to clarify the results,
we perform a quantum theoretically standard change of variables and set
\begin{eqnarray*}
E_0 &=&\frac{W}{2}\\
\Delta & =& \sqrt{\frac{1}{2}\Delta W^2 + 1 + \frac{1}{4}W^2 -W}\\
\tan\theta & =& \frac{2\sqrt{1+\frac{1}{4}W^2-W+\frac{1}{4}\Delta W^2}}{\Delta
  W}\\
e^{-\im\varphi} & =& \frac{1-\frac{1}{2}W - \im\frac{1}{2}\Delta W}{\sqrt{1 +
    \frac{1}{4}W^2 - W +\frac{1}{4}\Delta W^2}}.
\end{eqnarray*}
Then the eigenvalues are
\[
\lambda_1=E_0+\Delta, \quad \lambda_2=E_0-\Delta,
\]
and the unit eigenvectors take the form
\begin{equation}\label{eq:u}
u_1=\begin{pmatrix}\cos(\theta/2)
\\ e^{\im\varphi}\sin(\theta/2)\end{pmatrix}, \quad u_2
= \begin{pmatrix}-\sin(\theta/2)\\ e^{\im\varphi}\cos(\theta/2)\end{pmatrix}.
\end{equation}

Let us compute the evolution of the system by applying Schr\"odinger's equation,
taking the Hamiltonian to be the self-adjoint matrix $\hat{A}$, and  $\psi(0)=e_1=(1
\ \ 0)^T$.

$\psi(t)$ represents the state of the agents at timer $t$. Expressed in the $(u_1,u_2)$ basis, we find, with
the (reduced) Planck constant $\hbar = h/2\pi$,
\[
\psi(t) = \cos(\theta/2)e^{-\frac{\im}{\hbar}\lambda_1t}u_1
-\sin(\theta/2)e^{-\frac{\im}{\hbar}\lambda_2t}u_2
\]
and, in the standard $(e_1,e_2)$ basis, using (\ref{eq:u}):
\[
\psi(t) = e^{-\frac{\im}{\hbar}E_0t}\begin{pmatrix}\cos(t\Delta/\hbar)
  -\im\cos\theta\sin(t\Delta/\hbar)
  \\ -\im e^{\im\varphi}\sin(\theta)\sin(t\Delta/\hbar)\end{pmatrix}.
\]
The probability of transition from state $e_1=(1\ \ 0)^T$ to $e_2=(0\ \ 1)^T$
after time $t$ is given by
\[
\mathrm{Pr} = \|\langle e_2|\psi(t)\rangle\|^2 = \sin^2\theta\sin^2(t\Delta/\hbar).
\]
This probability has period $\pi \hbar/\Delta$ and maximal value
$\sin^2\theta$. In our case, we have
\[
\sin^2\theta =  \frac{\tan^2\theta}{\tan^2\theta+1}= 1 - \frac{\Delta W^2}{4 +
  W^2 - 4W + 2\Delta W^2},
\]
which is the maximal value of the probability of transition. Its period is
\[
\frac{\pi\hbar}{\sqrt{\frac{1}{2}\Delta W^2 + 1 + \frac{1}{4}W^2 -W}}.
\]

\medskip

Let us consider slightly different interaction matrices $A',A''$:
\[
A'=\begin{pmatrix} w_1 & w_1-1 \\ w_2-1 & w_2\end{pmatrix}, \quad
A''=\begin{pmatrix} w_1 & 1-w_1 \\ w_2-1 & w_2\end{pmatrix}
\]
with $w_1,w_2\in[0,1]$. In $A'$, the antidiagonal elements are nonpositive,
which, when $A'$ is interpreted as an influence matrix, corresponds to an
anticonformist attitude, since the agent has a tendency to adopt an opinion
opposite to the one of the other agent. $A''$ represents a mixed situation
where agent 1 is conformist and agent 2 is anticonformist. The corresponding
self-adjoint matrices are
\[
\widehat{A'} = \begin{pmatrix} w_1 & \frac{1}{2}W-1 +\im \frac{1}{2}\Delta W\\
\frac{1}{2}W-1 -\im \frac{1}{2}\Delta W & w_2\end{pmatrix}
\]
\[
\widehat{A''} = \begin{pmatrix} w_1 & \frac{1}{2}\Delta W +\im (\frac{1}{2} W-1)\\
\frac{1}{2}\Delta W -\im (\frac{1}{2} W-1) & w_2\end{pmatrix}
\]
Observe that since in all cases the term $A_{12}$ is the same, all parameters
$E_0,\Delta, \theta$ remain the same, only $\varphi$ changes. Consequently, the
eigenvalues and probability of transition are the same, only the evolution of
$\psi$ changes.

In the particular situation
\[
A=\begin{pmatrix}0 & 1 \\ 1 & 0\end{pmatrix},
\]
the agents display no activity (or no
self-confidence) and interact positively (or follow the other's opinion fully).
The eigenvalues are $\pm 1$, $\theta=\frac{\pi}{2}$ and $\Delta =1$,
yielding a maximal amplitude for probability transition and a period equal to
$\pi\hbar$. $e^{-\im\varphi}=1$ holds and the evolution is
\[
\psi(t) = \cos(t/\hbar)\begin{pmatrix}1\\0\end{pmatrix} - \im\sin(t/\hbar)\begin{pmatrix}0\\1\end{pmatrix}.
\]
Another extreme case is
\[
A=\begin{pmatrix}0 & -1 \\ -1 & 0\end{pmatrix},
\]
where the two agents have negative interaction or are pure anticonformists.
The eigenvalues are $\pm 1$, $\theta=\frac{\pi}{2}$ and $\Delta =1$,
yielding a maximal amplitude for probability transition and a period equal to
$\pi\hbar$. Again $e^{-\im\varphi}=-1$ holds with evolution
\[
\psi(t) = \cos(t/\hbar)\begin{pmatrix}1\\0\end{pmatrix} + \im\sin(t/\hbar)\begin{pmatrix}0\\1\end{pmatrix}.
\]
Note that in both cases, the state is oscillating with maximal amplitude, which
is well in line with the intuition that the opinion is very unstable in case every
agent just copies the activity/opinion of the other or does exactly the opposite.

\end{document}